\documentclass[12pt]{iopart}

\usepackage{graphicx}
\usepackage[utf8x]{inputenc}
\usepackage{color}
\usepackage{subfig}
\usepackage{multirow}
\usepackage{slashed}
\usepackage{cancel}
\usepackage{pdfsync}
\usepackage{booktabs}
\usepackage{float}
\usepackage{amssymb}
\usepackage{hyperref}

\begin{document}

\title[\small{The top quark chromomagnetic dipole moment in the SM from the 4-body vertex function}]{The top quark chromomagnetic dipole moment in the SM from the 4-body vertex function}

\author{J. Monta\~no-Dom\'inguez$^{1,2}$,
F. Ram\'irez-Zavaleta$^1$,
E. S. Tututi$^1$,
and
E. Urquiza-Trejo$^1$.
}
\vspace{10pt}
\address{$^1$Facultad de Ciencias F\'isico Matem\'aticas, Universidad Michoacana de San Nicol\'as de Hidalgo,
Av. Francisco J. M\'ugica s/n, 58060, Morelia, Michoac\'an, M\'exico.}
\address{$^2$CONAHCYT, Av. Insurgentes Sur 1582, Col. Cr\'edito Constructor, Alc. Benito Ju\'arez, 03940, CDMX, M\'exico.}
\ead{jmontano@conahcyt.mx}
\vspace{10pt}
\begin{indented}
\item[] October 2023
\end{indented}

\begin{abstract}
A new proposal to compute the anomalous chromomagnetic dipole moment of the top quark, $\hat{\mu}_t$, in the Standard Model is presented. On the basis of the 5-dimensional effective Lagrangian operator that characterizes the quantum-loop induced chromodipolar vertices $gt\bar{t}$ and $ggt\bar{t}$, the $\hat{\mu}_t$ anomaly is derived via radiative correction at the 1-loop level from the non-Abelian 4-body vertex function $ggt\bar{t}$. We evaluate $\hat{\mu}_t(s)$ as a function of the energy scale $s=\pm E^2$, for $E=[10,1000]$ GeV, taking into account the running of the quark masses and alpha strong through the $\overline{\mathrm{MS}}$ scheme. In particular, we find that at the typical energy scale $E=m_Z$ for high-energy physics, similarly to $\alpha_s(m_Z^2)$, $\alpha(m_Z^2)$ and $s_W(m_Z^2)$, the spacelike evaluation yields $\hat{\mu}_t(-m_Z^2)$ $=$ $-0.025$$+$$0.00384i$ and the timelike $\hat{\mu}_t(m_Z^2)$ $=$ $-0.0318$$-$$0.0106i$. This Re$\thinspace\hat{\mu}_t(-m_Z^2)$ $=$ $-0.025$ from $ggt\bar{t}$ is even closer to the experimental central value $\hat{\mu}_t^\mathrm{Exp}=$ $-0.024$, than that coming from the known 3-body vertex function $gt\bar{t}$, $-0.0224$. Once again, the Im$\thinspace\hat{\mu}_t(-m_Z^2)$ part is due to the contribution of virtual charged currents, just like in the $gt\bar{t}$ case. We can infer that the spacelike prediction is the favored one.
\end{abstract}

\begin{center}
\small{(This preprint matches with the published version.)}
\end{center}

%
%
%
%
%

\section{Introduction}
\label{sec:intro}

In the Standard Model (SM) the anomalous chromomagnetic dipole moment (CMDM), $\hat{\mu}_q$, is induced at the 1-loop level and receives virtual loop contributions from quantum chromodynamics (QCD), electroweak (EW) and Yukawa (YK) sectors~\cite{Choudhury:2014lna,Aranda:2018zis,Aranda:2021,Tavares,Montano-Dominguez:2021eeg}. Moreover, it has already been established in the literature that the non-static $\hat{\mu}_q$ is consistent with the properties of an observable, since this one is gauge invariant, gauge independent, infrared (IR) and ultraviolet (UV) finite~\cite{Aranda:2021,Tavares,Montano-Dominguez:2021eeg}. These CMDM results have only been derived from studies based on the 1-loop level radiative correction of the 3-body vertex $gq\bar{q}$. On the other hand, the 5-dimensional effective Lagrangian operator that characterizes $\hat{\mu}_q$ and the chromoelectric dipole moment (CEDM), $\hat{d}_q$, establishes that they are proportional to both the 3-body $gq\bar{q}$ and to the 4-body $ggq\bar{q}$ vertices~\cite{Haberl:1995ek}. Therefore, it is now interesting to obtain and analyze this observable that underlies from the $ggq\bar{q}$ coupling, by comparing it with the one coming from $gq\bar{q}$. Thus, this new phenomenological prediction will allow us to test the scope and power of the SM.

Regarding to the top quark CMDM, the experimental CMS Collaboration reported that $\hat{\mu}_t^\mathrm{Exp}=$ $-0.024_{-0.009}^{+0.013}(\mathrm{stat})$ $_{-0.011}^{+0.016}(\mathrm{syst})$~\cite{Sirunyan:2019eyu}, using proton-proton collisions at the centre-of-mass energy of 13 TeV with an integrated luminosity of 35.9 fb$^{-1}$. On the other hand, within the SM context, $\hat{\mu}_t$ has been exhaustively studied from the 3-body vertex $gt\bar{t}$  at the 1-loop level \cite{Choudhury:2014lna,Aranda:2018zis,Aranda:2021,Tavares}, where the most promising prediction establishes that $\mathrm{Re}\thinspace\hat{\mu}_t(-m_Z^2)=-0.0224$~\cite{Aranda:2018zis,Aranda:2021}, which supports the solidity of the model. This last result was evaluated in the spacelike domain ($s=-m_Z^2$), just as occurs for $\alpha_s(m_Z^2)$, $\alpha(m_Z^2)$ and $s_W(m_Z^2)$ which are conventionally fixed at the $Z$ gauge boson mass energy scale for high-energy physics processes~\cite{Workman:2022ynf,Field:1989uq,Deur:2016tte,dEnterria:2022hzv,Yndurain:2006amm,Nesterenko:2016pmx}. In addition, it was found an absorptive contribution, $\mathrm{Im}\thinspace\hat{\mu}_t(-m_Z^2)=-0.000925i$, which is strictly induced by virtual charged currents. In contrast, the timelike prediction was $\hat{\mu}_t(m_Z^2)=$ $-0.0133$ $-0.0267i$, whose real part deviates from the experimental measurement.

As it is seen from the 3-body vertex $gt\bar{t}$, a
slight discrepancy is still present between the experimental measurement and the theoretical prediction of the SM \cite{Aranda:2018zis,Aranda:2021,Tavares}. Motivated by this fact, starting from the 5-dimensional chromoelectromagnetic effective Lagrangian \cite{Haberl:1995ek} that predicts $\hat{\mu}_t$ proportional to the non-Abelian 4-body vertex, we propose to calculate $\hat{\mu}_t$ in the quantum field theory context as a result of virtual loop fluctuation of the vertex function $ggt\bar{t}$, that is to say, the chromodipole must also be induced from the radiative correction of such vertex at the 1-loop level, in a similar way as it is usually obtained via the 3-body vertex function $gt\bar{t}$ \cite{Choudhury:2014lna,Aranda:2018zis,Aranda:2021,Tavares,Bermudez:2017bpx}. However, the degree of difficulty to achive this is much higher; we will see that, once again and as expected, an infrared divergence emerges for the static case \cite{Choudhury:2014lna,Aranda:2018zis,Aranda:2021,Tavares,Bermudez:2017bpx,Eichten:1990vp}, which is a consequence of the non-Abelian nature of QCD.

It is very important to clarify, to avoid confusion, that the way of obtaining our $\hat{\mu}_t$ is process independent, it is not based on any scattering configuration such as $gg\to t\bar{t}$, $t\bar{t}\to gg$ or $gt\to gt$. Our procedure consists of calculating the $ggt\bar{t}$ vertex function as a 1-loop induced Feynman rule, from which we extract the Dirac structure $\sigma^{\mu\nu}$ (or $\gamma^\mu\gamma^\nu$) that will reveal $\hat{\mu}_t$.

The paper is organized as follows. In Sec.~\ref{Sec:CMDM-lagrangian} the effective Lagrangian that predicts the chromoelectromagnetic dipole moments is presented. In Sec.~\ref{Sec:CMDM-diagrams} an overview of the 1-loop calculation of the $ggt\bar{t}$ vertex to obtain the CMDM. In Sec.~\ref{Sec:CEDM} the vanishing of the CEDM is explicitly shown. In Sec.~\ref{Sec:results} the numerical results of the top quark CMDM, arising from the the 4-body coupling, are commented. Finally, Sec.~\ref{Sec:conclusions} addresses the conclusion of our work. In \ref{Appendix:IR-PaVes} infrared divergent scalar functions are explicitly shown, and in \ref{Appendix-input-values} the running of the quark masses in the $\overline{\mathrm{MS}}$ scheme is described.

\section{The chromomagnetic dipole moment}
\label{Sec:CMDM-lagrangian}

The 5-dimensional effective Lagrangian operator that characterizes the quantum-loop induced chromoelectromagnetic dipole moments (CEMDM) \cite{Haberl:1995ek,Bernreuther:2013aga,Khachatryan:2016xws}, is given by
\begin{eqnarray}\label{Effective-Lagrangian}
\mathcal{L}_\mathrm{eff} &=&-\frac{1}{2}\bar{q}_{A}\sigma^{\mu\nu}\left(\mu_q+id_q\gamma^5\right)q_{B}G_{\mu\nu}^aT_{AB}^a,
\end{eqnarray}
with the gluon strength field
\begin{equation}\label{}
G_{\mu\nu}^a=\partial_\mu g_\nu^a-\partial_\nu g_\mu^a-g_sf_{abc}g_\mu^bg_\nu^c,
\end{equation}
where $\mu_q$ is the chromomagnetic dipole moment and $d_q$ is the chromoelectric one, $\sigma^{\mu\nu}\equiv \frac{i}{2}[\gamma^\mu,\gamma^\nu]$, $\bar{q}_A$ and $q_B$ are spinor fields with quark color indices $A$ and $B$, $g_\mu^a$ is the gluon field with $a=1,...,8$, $T_{AB}^a$ are the color generators and $f_{abc}$ are the structure constants of the SU(3)$_C$ group; $g_s=\sqrt{4\pi\alpha_s}$ is the QCD group coupling constant, being $\alpha_s(m_Z^2)=0.1179$
the strong running coupling constant \cite{Workman:2022ynf,Field:1989uq,Deur:2016tte,dEnterria:2022hzv,Yndurain:2006amm,Nesterenko:2016pmx} stablished in the spacelike regime at the $Z$ gauge boson mass scale for high-energy physics phenomena.

It should be recalled that the CMDM coming from the Abelian 3-body vertex $gq\bar{q}$ has analogy with the anomalous magnetic dipole moment in quantum electrodynamics (QED), characterized also by a 
5-dimensional effective operator \cite{Roberts:2009xnh,Jegerlehner:2017gek}, this was calculated by Schwinger in 1948 from a radiative correction at the 1-loop level which led to the famous result $a_e=\alpha/(2\pi)$ \cite{Schwinger:1948iu,Schwinger:1949ra} (see also the modern reviews \cite{Roberts:2009xnh,Jegerlehner:2017gek}).

On the other hand, in contrast to QED, for QCD this situation is more sophisticated due to its non-Abelian nature which consequently predicts that the chromodipoles originate separately from two vertices: the Lagrangian (\ref{Effective-Lagrangian}) states that $\mu_q$ and $d_q$ are proportional to both the Abelian term $\partial_\mu g_\nu^a-\partial_\nu g_\mu^a$, responsible for the 3-body vertex $gq\bar{q}$, and to the non-Abelian term $g_sf_{abc}g_\mu^bg_\nu^c$, responsible for the 4-body vertex $ggq\bar{q}$. This is, the chromodipoles predicted in the 5-dimensional operator can be derived from both $gq\bar{q}$ and $ggq\bar{q}$ vertex functions, which in quantum field theory arise perturbatively as a result of virtual loop effects.

In this work we obtain $\mu_q$ from the non-Abelian 4-body vertex in the SM, in particular for the top quark, which numerically will reproduce almost faithfully the one corresponding to the Abelian 3-body vertex, while $d_q=0$ for both cases.

From (\ref{Effective-Lagrangian}), the Feynman rules for the chromodipolar interactions $gq\bar{q}$ and $ggq\bar{q}$ are
\begin{equation}\label{3b-vertex}
\Gamma_\mathrm{3b}^\mu=T_{AB}^a\sigma^{\mu\nu}q_\nu\left(\mu_q+id_q\gamma^5\right)
\end{equation}
and
\begin{equation}\label{4b-vertex}
\Gamma_\mathrm{4b}^{\mu\nu}=ig_sf_{abc}T^c_{AB}\sigma^{\mu\nu}\left(\mu_q+id_q\gamma^5\right),
\end{equation}
respectively, where $q_\nu$ is the gluon transfer momentum. In Fig.~\ref{FIGURE-CEMDM-loop-vertices} the quantum-loop induced Feynman rules are shown, where the quark momenta are defined as incoming for $q_B$ and outgoing for $q_A$, the gluon momenta are incoming. For $ggq\bar{q}$ the Bose Symmetry of the gluons has been considered.

According to the literature and for comparison with the experimental reports, it is convenient to define the dimensionless chromodipoles as~\cite{Haberl:1995ek,Workman:2022ynf,Bernreuther:2013aga,Khachatryan:2016xws}
\begin{equation}\label{}
\hat{\mu}_{q}\equiv \frac{m_q}{g_s}\mu_q \quad \mathrm{and} \quad \hat{d}_q\equiv \frac{m_q}{g_s}d_q ,
\end{equation}
where $m_q$ is the quark mass. The chromodipoles are in general complex quantities that can develop absorptive imaginary parts \cite{Bernreuther:2013aga,Khachatryan:2016xws}.

\begin{center}
\centering
\begin{figure}[t!]
\subfloat[]{\includegraphics[width=4.50cm]{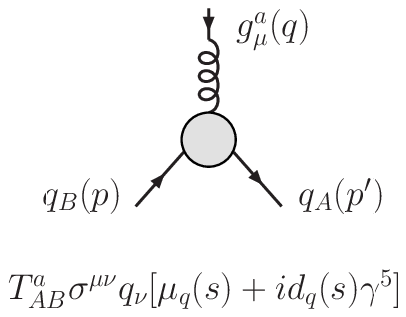}} \qquad \qquad
\subfloat[]{\includegraphics[width=5.25cm]{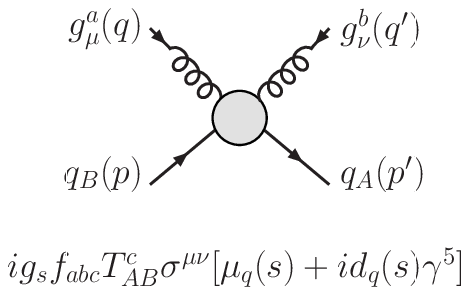}}
\caption{The quantum-loop induced CEMDM Feynman rules as function of the Lorentz-invariant scale energy $s$: (a) Abelian 3-body vertex function $gq\bar{q}$, with kinematics $q+p=p'$, and (b) non-Abelian 4-body vertex function $ggq\bar{q}$, with $q+q'+p=p'$.}
\label{FIGURE-CEMDM-loop-vertices}
\end{figure}
\end{center}

\section{The anomalous CMDM in the SM from the 1-loop vertex function $ggt\bar{t}$}
\label{Sec:CMDM-diagrams}

As discussed in the previous section, the CEMDM predicted in (\ref{Effective-Lagrangian}) are quantum-loop induced. Specifically, the chromodipoles will arise from both the Abelian 3-body $gt\bar{t}$ and the non-Abelian 4-body $ggt\bar{t}$ vertex functions at the 1-loop level. A detailed calculation of the chromodipoles from the 3-body vertex in the SM can be consulted in \cite{Aranda:2021}.

To distinguish the CEMDM derived from the radiative correction of the 4-body vertex we will denote them as $\hat{\mu}_t^\mathrm{4b}$ and $\hat{d}_t^\mathrm{4b}$, in order to compare them later with $\hat{\mu}_t^\mathrm{3b}$ and $\hat{d}_t^\mathrm{3b}$ belonging to the 3-body vertex \cite{Aranda:2021}. In advance, it will result $\hat{d}_t^\mathrm{4b}=0$ in the SM, just as it happened with $\hat{d}_t^\mathrm{3b}=0$, for this reason we will focus the discussion on $\hat{\mu}_t^\mathrm{4b}$.

As it concerns to the calculation of $\hat{\mu}_t^\mathrm{3b}(s)$ from the 3-body vertex function, with configuration $g_\mu^a(q)t(p)\bar{t}(p')$ (Fig.~\ref{FIGURE-CEMDM-loop-vertices}a) and kinematics $q+p=p'$, it has only one Lorentz-invariant momentum transfer $q^2=(p'-p)^2\equiv s$ that characterizes its energy scale dependence. In this case $q^2$ coincides with the Lorentz-invariant $s$ of the vertex. It should be recalled that the Lorentz-invariant $s$ can not be static, this is, it must be different from zero in order to avoid an IR divergence, that is why the gluon must be off-shell, $s\equiv q^2\neq 0$; this particular behaviour is originated from the Feynman loop diagram that includes the self-interacting non-Abelian vertex $ggg$ (diagram Fig.~2f in \cite{Aranda:2021}).

On the other hand, in Fig.~\ref{FIGURE-CEMDM-loop-vertices}b we set the configuration $g_\mu^a(q)g_\nu^b(q')t(p)\bar{t}(p')$
for the 4-body vertex function, with kinematics $q+q'+p=p'$, $q^2=q'^{\thinspace 2}=0$ and $p^2=p'^{\thinspace 2}=m_{t}^2$. Unlike the $gt\bar{t}$ vertex, $ggt\bar{t}$ is not restricted to have the gluons on-shell. The topology of $ggt\bar{t}$ generates several Lorentz-invariant momentum transfer configurations, $(q+q')^2$ $=$ $(p-p')^2$, $(q+p)^2$ $=$ $(p'-q')^2$ and $(p+q')^2$ $=$ $(p'-q)^2$. We will assume them at the same energy scale $s$, because in the vertex function none of them is privileged over any other, that is to say, all are equally important: $(q+q')^2$ $=$ $(p-p')^2$ $=$ $(q+p)^2$ $=$ $(p'-q')^2$ $=$ $(p+q')^2$ $=$ $(p'-q)^2$ $\equiv$ $s\neq 0$. If $s=0$ then $\hat{\mu}_t^\mathrm{4b}(0)$ develops an IR divergence, as in  $\hat{\mu}_t^\mathrm{3b}(0)$ \cite{Aranda:2021}.

The invariant amplitude of the whole $ggt\bar{t}$ vertex function can be expressed as
\begin{equation}\label{}
\mathcal{M}=\mathcal{M}^{\mu\nu} \epsilon_\mu^a(\vec{q}\thinspace)\epsilon_\nu^b(\vec{q}\thinspace'),
\end{equation}
where $\epsilon_\mu^a(\vec{q}\thinspace)$ and $\epsilon_\nu^b(\vec{q}\thinspace')$ are the polarization vectors of the gluons, being the tensor amplitude written as
\begin{equation}\label{general-amplitude}
\mathcal{M}^{\mu\nu}=\bar{u}(p')\Gamma^{\mu\nu}u(p),
\end{equation}
were $\bar{u}(p')$ and $u(p)$ are the quark spinors, and $\Gamma^{\mu\nu}$ is the Lorentz structure of the whole vertex, from which we should extract the chromodipole term of interest $\Gamma_\mathrm{4b}^{\mu\nu}$, given in Eq.~(\ref{4b-vertex}).

\begin{center}
\begin{figure}[t!]
\centering
\includegraphics[scale=0.925]{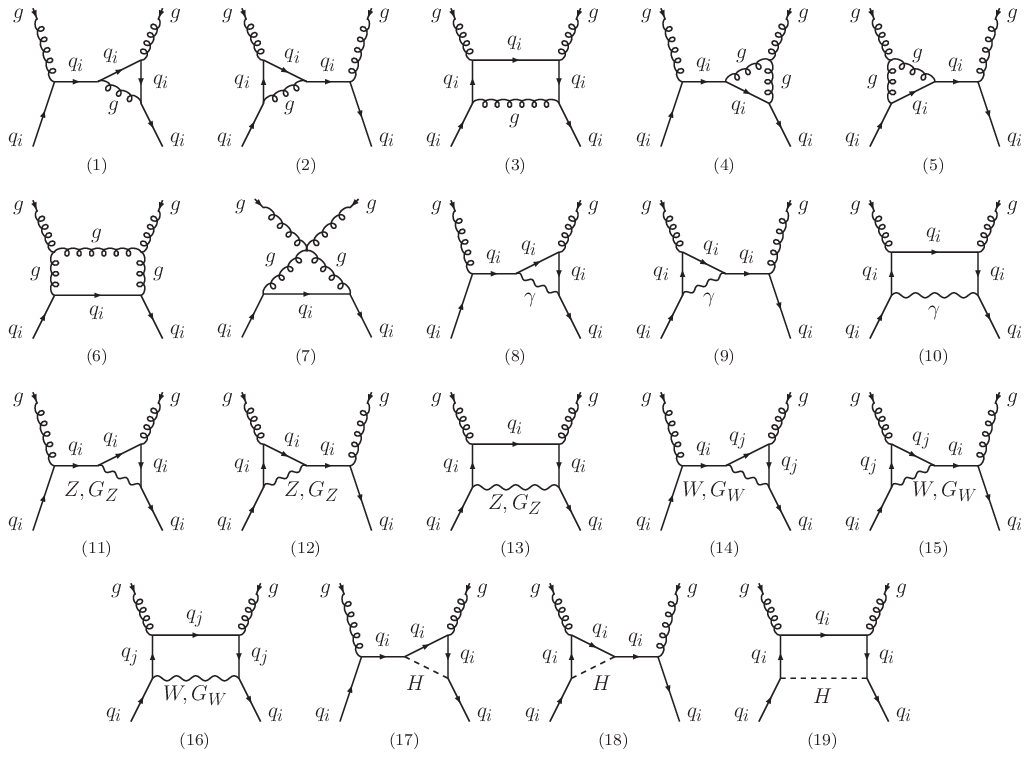}
\caption{1-Loop diagrams that induce the CMDM from the non-Abelian 4-body vertex in the $\xi=1$ Feynman-`t Hooft gauge, where $q_i=t$ and $q_j=d,s,b$; in total there are 73 diagrams due to the Bose symmetry of external gluons.}
\label{FIGURE-Diagrams-4B}
\end{figure}
\end{center}

In the SM, the 1-loop diagrams that give rise to $\hat{\mu}_{t}^\mathrm{4b}(s)$ are displayed and condensed in Fig.~\ref{FIGURE-Diagrams-4B}. The calculation is carried out in the $\xi=1$ Feynman$-$`t Hooft gauge, the Bose symmetry of the external gluons is implied, involving up to 73 diagrams. These contributions were generated by \texttt{FeynArts}~\cite{Hahn:2000kx} and solved with \texttt{FeynCalc}~\cite{Mertig:1990an,Shtabovenko:2016sxi,Shtabovenko:2020gxv}, \texttt{ColorMath} \cite{Sjodahl:2012nk} was employed to address the color algebra.

The amplitude can be split out into five sets distinguished by the different virtual bosons $g$, $\gamma$, $Z$, $W$ and $H$:
\begin{equation}\label{Amplitude-4b}
\mathcal{M}=\mathcal{M}(g)+\mathcal{M}(\gamma)
+\mathcal{M}(Z)+\mathcal{M}(W)+\mathcal{M}(H),
\end{equation}
where in Fig.~\ref{FIGURE-Diagrams-4B} the $G_Z$ and $G_W$ are pseudo-Goldstone bosons of the $Z$ and $W$ gauge bosons, respectively.

In general, the resulting tensor amplitudes produce many different Dirac terms, where only the two terms $\gamma^\mu\gamma^\nu$ and $\gamma^\mu\gamma^\nu\gamma^5$, proportional to $f_{abc}T^c$, will lead to identify the CEMDM. The $\gamma^5$ term comes from the axial vector components of the $Z$ and $W$ contributions. Thus, the general tensor amplitude that contains the terms and form factors of interest can be written as
\begin{eqnarray}\label{general-amplitude}
\mathcal{M}^{\mu\nu} &=& \overline{u}(p')
f_{abc}T^c \left(f_1\gamma^\mu\gamma^\nu+f_2\gamma^\mu\gamma^\nu\gamma^5+...\right)
u(p)
\nonumber\\
&=& \overline{u}(p')
f_{abc}T^c \left(-if_1\sigma^{\mu\nu}-if_2\sigma^{\mu\nu}\gamma^5+...\right)
u(p),
\end{eqnarray}
where we have used $\gamma^\mu\gamma^\nu=g^{\mu\nu}\textbf{1}-i\sigma^{\mu\nu}$ from $\{\gamma^\mu,\gamma^\nu\}=2g^{\mu\nu}\textbf{1}$, and the SU(3)$_C$ generator properties $[T^x,T^y]=$ $if_{xyz}T^z$ and $T^xT^y=$ $1/2\big[1/N_C\delta_{xy}\textbf{1}$ $+(d_{xyz}$ $+if_{xyz})T^z\big]$, with $N_C=3$. From (\ref{general-amplitude}) we isolate and identify the desired CEMDM subamplitude by comparing it with the reference vertex $\Gamma^{\mu\nu}_\mathrm{4b}$ in Eq.~(\ref{4b-vertex}), thus
\begin{eqnarray}\label{4b-amplitude}
\mathcal{M}_\mathrm{CEMDM}^{\mu\nu} & =ig_sf^{abc}T^a\sigma^{\mu\nu}\left[\left(-\frac{1}{g_s}f_1\right)+i\left(\frac{i}{g_s}f_2\right)\gamma^5\right]
\nonumber\\
& =ig_sf_{abc}T^c_{AB}\sigma^{\mu\nu}\left(\mu_t+id_t\gamma^5\right),
\end{eqnarray}
where
\begin{equation}\label{}
\hat{\mu}_{t}=-\frac{m_{t}}{g_s^2}f_1 \quad \mathrm{and} \quad \hat{d}_{t}=\frac{im_{t}}{g_s^2}f_2=0.
\end{equation}
The general $\hat{\mu}_{t}$ is UV and IR finite in all the virtual contributions $\gamma$, $Z$, $W$, and $H$, except in the $g$ case where an IR divergence triggers if the Lorentz-invariant $s=0$; while $\hat{d}_{t}=0$, because each $Z$ and $W$ contribution add up to zero, we show this in Sec.~\ref{Sec:CEDM}.

According to the amplitude in Eq.~(\ref{Amplitude-4b}), $\hat{\mu}_{t}$ is composed by five sets of contributing virtual particles (see Fig. \ref{FIGURE-Diagrams-4B}) as
\begin{equation}\label{CMDM-4b}
\hat{\mu}_{t}^\mathrm{4b}(s)=
\hat{\mu}_{t}(g)+\hat{\mu}_{t}(\gamma)+\hat{\mu}_{t}(Z)+\hat{\mu}_{t}(W)+\hat{\mu}_{t}(H),
\end{equation}
where the Lorentz-invariant momentum transfer of the $ggt\bar{t}$ vertex must obey $s\neq 0$.

Thus, 73 diagrams contribute to the $ggt\bar{t}$ vertex. It should be recalled that if the Lorentz invariants are fixed at $s=0$, then $\hat{\mu}_t^\mathrm{4b}(0)$ develops an IR divergence, this behavior comes exclusively from the (6) and (7) diagrams, which involves self-interacting non-Abelian vertices $ggg$ and $gggg$, respectively, generating three and two virtual gluons inside the loops.

Below we explicitly present the five different sets of virtual-contributing particles to $\hat{\mu}_t^\mathrm{4b}(s)$.

\subsection{The $g$ contribution}
\label{g-contribution}

The gluon contribution $\hat{\mu}_{t}(g)$ comprises the first seven loops $(1)-(7)$ in Fig. \ref{FIGURE-Diagrams-4B}, although there are actually 13 diagrams, since the $(1)-(6)$ set must consider the Bose symmetry (BS) for external gluons; this symmetry is achieved by exchanging $(a,\mu,q)$ $\leftrightarrow$ $(b,\nu,q')$ where these gluons (indistinguishable particles) income in different diagram points.

This gluon contribution is the most complicated case. We have constructed by hand each loop diagram, and solve its several non-trivial color algebras involved. The result was corroborated by using three different procedures in \texttt{Mathematica} 11:
i) tensor loop integrals solved with \texttt{FeynCalc} 8.2 \cite{Mertig:1990an} in 4 dimensions, and the Passarino-Veltman scalar functions (PaVes) analytically and numerically solved with \texttt{Package-X} 2.1.1 \cite{Patel:2015tea};
ii) tensor loop integrals solved with \texttt{FeynCalc} 10 in D dimensions \cite{Shtabovenko:2016sxi,Shtabovenko:2020gxv}, and the PaVes solved with \texttt{Package-X};
iii) tensor loop integrals solved entirely analytically and numerically with pure \texttt{Package-X}.
Such procedures lead us exactly to the same $\hat{\mu}_{t}(g)$ numerical result.

Aditionally, in an automated way, with $\texttt{FeynArts}$ \cite{Hahn:2000kx} we crossed-check each loop diagram amplitude by using two different ways: i) loop integrals generated in 4 dimensions and solved with \texttt{FeynCalc} 8.2, and ii) loop integrals in D dimensions and solved with \texttt{FeynCalc} 10. The outcomes in terms of the PaVes were solved with \texttt{Package-X}. These automated procedures confirmed faithfully our handmade calculations.

Thus, the $\hat{\mu}_{t}(g)$ contribution with $s\neq 0$, from \texttt{FeynCalc} 8.2 in 4 dimensions, is
\begin{eqnarray}\label{CMDM-g}
\hat{\mu}_t(g)&= \frac{\alpha_s m_t^2}{24\pi(4m_t^2-s)}\Bigg[
-34\frac{\left(5 m_t^2-2 s\right) \left(10 m_t^4-18 m_t^2 s+5 s^2\right)}{\left(m_t^2-s\right) \left(9 m_t^4-16 m_t^2 s+4 s^2\right)} B_{0(1)}^g
\nonumber\\
& +72 B_{0(2)}^g
+34\frac{ \left(4 m_t^2-s\right) \left(12 m_t^4-23 m_t^2 s+8 s^2\right)}{\left(m_t^2-s\right) \left(9 m_t^4-16 m_t^2 s+4 s^2\right)}B_{0(3)}^g
-4 B_{0(4)}^g
\nonumber\\
&-72\frac{\left(2 m_t^2-3 s\right) \left(4 m_t^2-s\right)}{9 m_t^4-16 m_t^2 s+4 s^2}B_{0(5)}^g
+4\frac{\left(4 m_t^2-s\right) \left(2 m_t^2-3 s\right)}{9 m_t^4-16 m_t^2 s+4 s^2}B_{0(6)}^g
\nonumber\\
&+18\left(2 m_t^2-s\right)C_{0(1)}^g
-4\left(m_t^2-s\right)C_{0(2)}^g
+2\left(m_t^2-s\right)C_{0(3)}^g
\nonumber\\
&-9\left(2 m_t^2-s\right)C_{0(4)}^g
+36sC_{0(5)}^g
+2\left(4 m_t^2-s\right)C_{0(6)}^g
\nonumber\\
&+9\frac{14 m_t^6-79 m_t^4 s+100 m_t^2 s^2-20 s^3}{9 m_t^4-16 m_t^2 s+4 s^2}C_{0(7)}^g
\nonumber\\
&-2\frac{31 m_t^6-61 m_t^4 s+23 m_t^2 s^2-2 s^3}{9 m_t^4-16 m_t^2 s+4 s^2}C_{0(8)}^g
+9s\left(2 m_t^2-s\right)D_{0(1)}^g
\nonumber\\
&+2\left(m_t^2-s\right) \left(4 m_t^2-s\right)D_{0(2)}^g \Bigg],
\end{eqnarray}
where the PaVes are written according to the $\texttt{Package-X}$ notation:

$B_{0(1)}^g$ $\equiv$  $B_0(m_t^2; 0,m_t)$,

$B_{0(2)}^g$ $\equiv$  $B_0(s\thinspace;0,0)$,

$B_{0(3)}^g$ $\equiv$  $B_0(s\thinspace;0,m_t)$,

$B_{0(4)}^g$ $\equiv$  $B_0(s\thinspace;m_t,m_t)$,

$B_{0(5)}^g$ $\equiv$  $B_0(-2m_t^2+3s\thinspace;0,0)$,

$B_{0(6)}^g$ $\equiv$  $B_0(-2m_t^2+3s\thinspace;m_t,m_t)$,

$C_{0(1)}^g$ $\equiv$  $C_0(0,s,-2m_t^2+3s\thinspace;0,0,0)$,

$C_{0(2)}^g$ $\equiv$  $C_0(0,s,-2m_t^2+3s\thinspace;m_t,m_t,m_t)$,

$C_{0(3)}^g$ $\equiv$  $C_0(m_t^2,0,s\thinspace;0,m_t,m_t)$,

$C_{0(4)}^g$ $\equiv$  $C_0(m_t^2,0,s\thinspace;m_t,0,0)$,

$C_{0(5)}^g$ $\equiv$  $C_0(m_t^2,m_t^2,s\thinspace;0,m_t,0)$,

$C_{0(6)}^g$ $\equiv$  $C_0(m_t^2,m_t^2,s\thinspace;$ $m_t,0,m_t)$,

$C_{0(7)}^g$ $\equiv$  $C_0(m_t^2,s,-2m_t^2+3s\thinspace;0,m_t,0)$,

$C_{0(8)}^g$ $\equiv$  $C_0(m_t^2,s,-2m_t^2+3s\thinspace;$ $m_t,0,m_t)$,

$D_{0(1)}^g$ $\equiv$  $D_0(m_t^2,m_t^2,0,$ $-2m_t^2+3s,s,s\thinspace;0,m_t,0,0)$,

$D_{0(2)}^g$ $\equiv$  $D_0(m_t^2,m_t^2,0,$ $-2m_t^2+3s,s,s\thinspace;m_t,0,m_t,m_t)$.

It is worth to emphasize that $\hat{\mu}_{t}(g)$ is free of UV divergences, and it is also free of IR divergences as long as $s\neq$0. By using \texttt{Package-X} we can appreciate that the ordinary $B_0$'s cancel each other out its $1/\epsilon_\mathrm{UV}$ poles. However, some PaVes are individually IR divergent even though $s\neq 0$, namely, the $C_{0(1,4,6)}^g$ and $D_{0(1,2)}^g$. These IR divergent functions arise through the box diagrams (3) and (6) in Fig.~\ref{FIGURE-Diagrams-4B}, but its $1/\epsilon_\mathrm{IR}$ and $1/\epsilon_\mathrm{IR}^2$ poles are canceled after adding all together; their solutions are explicitly given in \ref{Appendix:IR-PaVes}. The remaining PaVes and the continued dilogarithm $\mathcal{L}i_2$ will be directly evaluated numerically with \texttt{Package-X}.

\subsection{The $\gamma$ contribution}
\label{A-contribution}

The photon contribution is formed by the set of diagrams $(8)-(10)$ in Fig.~\ref{FIGURE-Diagrams-4B}, which provides 6 diagrams in total due to BS, resulting in
\begin{eqnarray}
\hat{\mu}_t(\gamma) &=
\frac{\alpha\thinspace Q_t^2}{2\pi}\thinspace \frac{m_t^2}{4m_t^2-s}
\Bigg[-\frac{\left(5 m_t^2-2 s\right) \left(10 m_t^4-18 m_t^2 s+5 s^2\right)}{\left(m_t^2-s\right) \left(9 m_t^4-16 m_t^2 s+4 s^2\right)} B_{0(1)}^\gamma
\nonumber\\
&+\frac{\left(4 m_t^2-s\right) \left(12 m_t^4-23 m_t^2 s+8 s^2\right)}
{\left(m_t^2-s\right) \left(9 m_t^4-16 m_t^2 s+4 s^2\right)} B_{0(3)}^\gamma
+2 B_{0(4)}^\gamma
\nonumber\\
& -\frac{2\left(2 m_t^2-3 s\right) \left(4 m_t^2-s\right)}{9 m_t^4-16 m_t^2 s+4 s^2} B_{0(6)}
+2\left(m_t^2-s\right) C_{0(2)}^\gamma
+\left(s-m_t^2\right)C_{0(3)}^\gamma
\nonumber\\
& -\left(4 m_t^2-s\right) C_{0(6)}^\gamma
+\frac{31 m_t^6-61 m_t^4 s+23 m_t^2 s^2-2 s^3}{9 m_t^4-16 m_t^2 s+4 s^2} C_{0(8)}^\gamma
\nonumber\\
& -\left(m_t^2-s\right) \left(4 m_t^2-s\right) D_{0(2)}^\gamma \Bigg] ,
\end{eqnarray}
where the PaVes are labeled as in the gluon case. Note that the functions $C_{0(6)}^\gamma$ and $D_{0(2)}^\gamma$ are IR divergent, but their $1/\epsilon_\mathrm{IR}$ cancels each other, and to appreciate this let us focus on the IR pieces from $\hat{\mu}_t(\gamma)$ (see the IR PaVes solutions in \ref{Appendix:IR-PaVes}):
\begin{eqnarray}
& C_{0(6)\mathrm{IR}}^\gamma 
+\left(m_t^2-s\right) D_{0(2)\mathrm{IR}}^\gamma
\nonumber\\
& = \frac{-1}{4 m_{t}^2-s}
\left(\Delta_\mathrm{IR}+\ln\frac{\mu ^2}{m_{t}^2}\right)
\frac{R_1}{s} \ln \frac{2 m_t^2-s+R_1}{2m_t^2}
\nonumber\\
& +\left(m_t^2-s\right)
\left[
\frac{1}{(m_t^2-s)(4 m_{t}^2-s)}
\left(\Delta_\mathrm{IR}+\ln\frac{\mu ^2}{m_{t}^2}\right)
\frac{R_1}{s} \ln \frac{2 m_t^2-s+R_1}{2m_t^2}
\right]
\nonumber\\
& =0 ,
\end{eqnarray}
therefore $\hat{\mu}_t(\gamma)$ is finite.

\subsection{The $Z$ contribution}
\label{Z-contribution}

The virtual $Z$ gauge boson contribution is composed by the $(11)-(13)$ diagrams set in Fig.~\ref{FIGURE-Diagrams-4B}, and due to the use of the Feynman$-$`t Hooft gauge $\xi=1$ it will appear a virtual replica contribution of its unphysical $G_Z$ pseudo-Goldstone boson, and because of the BS in total there are 12 diagrams. Thus, the final full $Z$ gauge boson contribution can be summarized in just 6 diagrams that already imply the sum of both virtual contribution parts: the physical $Z$ plus the unphysical $G_Z$. Therefore, the complete tensor amplitude of the $Z$ gauge boson contribution can be expressed as
\begin{eqnarray}
\mathcal{M}_t^{\mu\nu}(Z) &=
\sum_{i=11}^{13} \left[\mathcal{M}_{t\thinspace i}^{\mu\nu}(Z)_{\xi=1}+\mathcal{M}_{t\thinspace i}^{\mu\nu}(G_Z)_{\xi=1}\right]+\mathrm{BS}
\nonumber\\
&= \sum_{i=11}^{13} \mathcal{M}_{t\thinspace i}^{\mu\nu}(Z)+\mathrm{BS}
\nonumber\\
&= 
 \mathcal{M}_{t\thinspace 11}^{\mu\nu}
+\mathcal{M}_{t\thinspace 12}^{\mu\nu}
+\mathcal{M}_{t\thinspace 13}^{\mu\nu}
+\mathcal{M}_{t\thinspace 11\mathrm{BS}}^{\mu\nu}
+\mathcal{M}_{t\thinspace 12\mathrm{BS}}^{\mu\nu}
+\mathcal{M}_{t\thinspace 13\mathrm{BS}}^{\mu\nu},
\end{eqnarray}
in consequence
\begin{eqnarray}\label{Z-CMDM-parts}
 \hat{\mu}_t(Z) &=
 \hat{\mu}_{t\thinspace 11}
+\hat{\mu}_{t\thinspace 12}
+\hat{\mu}_{t\thinspace 13}
+\hat{\mu}_{t\thinspace 11\mathrm{BS}}
+\hat{\mu}_{t\thinspace 12\mathrm{BS}}
+\hat{\mu}_{t\thinspace 13\mathrm{BS}}.
\end{eqnarray}

The resulting final full contribution is
\begin{eqnarray}\label{}
\hat{\mu}_t(Z) &= \frac{\alpha}{8 \pi c_W^2s_W^2}\thinspace
\frac{m_t^2}{4 m_t^2-s}
\Bigg\{
\frac{g_{Vt}^2 m_Z^2+g_{At}^2 \left(m_Z^2-2m_t^2\right)}{m_Z^2}
\nonumber\\
&\times\bigg[-\frac{\left(5 m_t^2-2 s\right) \left(10 m_t^4-18 m_t^2 s+5 s^2\right)}
{\left(m_t^2-s\right) \left(9 m_t^4-16 m_t^2 s+4 s^2\right)} B_{0(1)}^Z
+2 B_{0(2)}^Z
\nonumber\\
&+\frac{\left(4 m_t^2-s\right) \left(12 m_t^4-23 m_t^2 s+8 s^2\right)}
{\left(m_t^2-s\right) \left(9 m_t^4-16 m_t^2 s+4 s^2\right)} B_{0(3)}^Z
\nonumber\\
& -2\frac{\left(2 m_t^2-3 s\right) \left(4 m_t^2-s\right)}{\left(9 m_t^4-16 m_t^2 s+4 s^2\right)} B_{0(4)}^Z
+2\left(m_t^2-s\right) C_{0(1)}^Z
\bigg]
\nonumber\\
& +\left[g_{Vt}^2 \left(s-m_t^2\right)
+g_{At}^2\frac{m_t^2\left(10 m_t^2-m_Z^2-4 s\right)+m_Z^2 s}{m_Z^2}\right] C_{0(2)}^Z
\nonumber\\
&+\big[g_{Vt}^2 \left(-4 m_t^2+2 m_Z^2+s\right)
+g_{At}^2 \left(-8 m_t^2+2 m_Z^2+s\right)\big] C_{0(3)}^Z
\nonumber\\
&+\bigg[g_{Vt}^2\frac{31 m_t^6-m_t^4 \left(16 m_Z^2+61 s\right)+m_t^2 s \left(28 m_Z^2+23 s\right)-2 s^2 \left(3 m_Z^2+s\right)}{9 m_t^4-16 m_t^2 s+4 s^2}
\nonumber\\
& +g_{At}^2 \bigg(\frac{m_t^2}{m_Z^2}\thinspace
\frac{
10 m_t^6+3 m_t^4 \left(21 m_Z^2-8 s\right)
+m_t^2 \left(-16 m_Z^4-117 m_Z^2 s+18 s^2\right)
}{9 m_t^4-16 m_t^2 s+4 s^2}
\nonumber\\
&+\frac{m_t^2}{m_Z^2}\thinspace
\frac{
s \left(28 m_Z^4+35 m_Z^2 s-4 s^2\right)
}{9 m_t^4-16 m_t^2 s+4 s^2}
-\frac{2s^2 \left(3 m_Z^2+s\right)}{9 m_t^4-16 m_t^2 s+4 s^2}\bigg)
\bigg]C_{0(4)}^Z
\nonumber\\
&-\left(m_t^2-s\right) \left[g_{Vt}^2 \left(4 m_t^2-2 m_Z^2-s\right)+g_{At}^2 \left(8 m_t^2-2 m_Z^2-s\right)\right] D_{0(1)}^Z \Bigg\},
\nonumber\\
\end{eqnarray}
with

$B_{0(1)}^Z$ $\equiv$ $B_0\left(m_t^2; m_t, m_Z\right)$,

$B_{0(2)}^Z$ $\equiv$ $B_0\left(s\thinspace; m_t m_t\right)$,

$B_{0(3)}^Z$ $\equiv$ $B_0\left(s\thinspace; m_t, m_Z\right)$,

$B_{0(4)}^Z$ $\equiv$ $B_0\left(-2m_t^2 + 3s\thinspace; m_t, m_t\right)$,

$C_{0(1)}^Z$ $\equiv$ $C_0\left(0, s, -2m_t^2 + 3s\thinspace; m_t, m_t, m_t\right)$,

$C_{0(2)}^Z$ $\equiv$ $C_0\left(m_t^2, 0, s\thinspace; m_Z, m_t, m_t\right)$,

$C_{0(3)}^Z$ $\equiv$ $C_0\left(m_t^2, m_t^2, s\thinspace; m_t, m_Z, m_t\right)$,

$C_{0(4)}^Z$ $\equiv$ $C_0\left(m_t^2, s, -2m_t^2 + 3s\thinspace; m_t, m_Z, m_t\right)$,

$D_{0(1)}^Z$ $\equiv$ $D_0\left(m_t^2, m_t^2, 0, -2m_t^2 + 3s, s, s\thinspace; m_t, m_Z, m_t, m_t\right)$.

\noindent All these $C_0$'s and $D_0$'s are finite and will be evaluated with \texttt{Package-X}; the same procedure applies for the $W$ and $H$ contributions.

\subsection{The $W$ contribution}
\label{W-contribution}

The $W$ gauge boson contribution is represented by the set of $(14)-(16)$ loops in Fig.~\ref{FIGURE-Diagrams-4B}, considering its unphysical $G_W$ pseudo-Goldstone boson replica contribution due to the gauge $\xi=1$ (as in the $Z$ case), and because of the BS there are 12 diagrams for each $q_j$ $=$ $q_1,q_2,q_3$ $=$ $d,s,b$ quark. The final full result, that already imply the sum of the virtual physical $W$ plus the unphysical $G_W$ contributions, is
\begin{eqnarray}
\hat{\mu}_t(W) &= \frac{\alpha}{16\pi}\thinspace\frac{m_t^2}{m_W^2}
\sum_{j=1}^3 \|V_{tq_j}\|^2
\frac{m_t^2+m_{q_j}^2-2 m_W^2}{4 m_t^2-s}
\nonumber\\
&\times
\Bigg(
\frac{\left(5 m_t^2-2 s\right) \left(10 m_t^4-18 m_t^2 s+5 s^2\right)}{\left(m_t^2-s\right) \left(9 m_t^4-16 m_t^2 s+4 s^2\right)} B_{0(1)}^W
-2 B_{0(2)}^W
\nonumber\\
& -\frac{\left(4 m_t^2-s\right) \left(12 m_t^4-23 m_t^2 s+8 s^2\right)}{\left(m_t^2-s\right) \left(9 m_t^4-16 m_t^2 s+4 s^2\right)} B_{0(3)}^W
\nonumber\\
& +2\frac{\left(2 m_t^2-3 s\right) \left(4 m_t^2-s\right)}{9 m_t^4-16 m_t^2 s+4 s^2} B_{0(4)}^W
-2 \left(m_t^2-s\right) C_{0(1)}^W
\nonumber\\
& +\frac{m_t^4+m_t^2 \left(9 m_{q_j}^2-2 m_W^2-s\right)-3 m_{q_j}^2 s+2 m_W^2 s}{m_t^2+m_{q_j}^2-2m_W^2} C_{0(2)}^W
\nonumber\\
&+\Bigg[\frac{2 m_t^4-m_t^2 \left(4 m_{q_j}^2+6 m_W^2+s\right)+2 m_{q_j}^4
}{m_t^2+m_{q_j}^2-2 m_W^2}
\nonumber\\
&+\frac{m_{q_j}^2 \left(s-6 m_W^2\right)+2 m_W^2 \left(2 m_W^2+s\right)}{m_t^2+m_{q_j}^2-2 m_W^2}
\Bigg] C_{0(3)}^W
\nonumber\\
& +\frac{1}{4} \Bigg\{
\frac{-9 m_t^4+m_t^2 \left[23 m_{q_j}^2+2 \left(9 m_W^2+s\right)\right]-2 s \left(3 m_{q_j}^2+2 m_W^2\right)}{m_t^2+m_{q_j}^2-2 m_W^2}
\nonumber\\
&+\frac{21 m_t^6+m_t^4 \left(-64 m_{q_j}^2+64 m_W^2-30 s\right)
}{9 m_t^4-16 m_t^2 s+4 s^2}
\nonumber\\
&+\frac{\left(m_{q_j}^2-m_W^2\right)
\left(112 m_t^2 s-24 s^2\right)}{9 m_t^4-16 m_t^2 s+4 s^2}
\Bigg\} C_{0(4)}^W
\nonumber\\
&+\frac{m_t^2-s}{m_t^2+m_{q_j}^2-2 m_W^2}
\Big[m_t^2\left(2 m_t^2-4 m_{q_j}^2-6 m_W^2-s\right)
\nonumber\\
&+m_{q_j}^2\left(2 m_{q_j}^2+s-6 m_W^2\right)+2 m_W^2 \left(2 m_W^2+s\right)\Big] D_{0(1)}^W
\Bigg),
\end{eqnarray}
with

$B_{0(1)}^W$ $\equiv$ $B_0\left(m_t^2;m_{q_j},m_W\right)$,

$B_{0(2)}^W$ $\equiv$ $B_0\left(s\thinspace;m_{q_j},m_{q_j}\right)$,

$B_{0(3)}^W$ $\equiv$ $B_0\left(s\thinspace;m_{q_j},m_W\right)$,

$B_{0(4)}^W$ $\equiv$ $B_0\left(-2m_t^2+3s\thinspace;m_{q_j},m_{q_j}\right)$,

$C_{0(1)}^W$ $\equiv$ $C_0\left(0,s,-2m_t^2+3s\thinspace;m_{q_j},m_{q_j},m_{q_j}\right)$,

$C_{0(2)}^W$ $\equiv$ $C_0\left(m_t^2,0,s\thinspace;m_W,m_{q_j},m_{q_j}\right)$,

$C_{0(3)}^W$ $\equiv$ $C_0\left(m_t^2,m_t^2,s\thinspace;m_{q_j},m_W,m_{q_j}\right)$,

$C_{0(4)}^W$ $\equiv$ $C_0\left(m_t^2,s,-2m_t^2+3s\thinspace;m_{q_j},m_W,m_{q_j}\right)$,

$D_{0(1)}^W$ $\equiv$ $D_0\left(m_t^2,m_t^2,0,-2m_t^2+3s,s,s\thinspace;m_{q_j},m_W,m_{q_j},m_{q_j}\right)$.

\subsection{The $H$ contribution}

The Higgs contribution encompasses the $(17)-(19)$ diagrams in Fig.~\ref{FIGURE-Diagrams-4B}, which in fact are 6 in total by BS, leading to
\begin{eqnarray}
\hat{\mu}_t(H) &= \frac{\alpha}{16\pi}\thinspace\frac{m_t^4}{m_W^2\left(4 m_t^2-s\right)}
\Bigg[\frac{\left(5 m_t^2-2 s\right) \left(10 m_t^4-18 m_t^2 s+5 s^2\right)}{\left(m_t^2-s\right) \left(9 m_t^4-16 m_t^2 s+4 s^2\right)} B_{0(1)}^H
\nonumber\\
& -2 B_{0(2)}^H
-\frac{\left(4 m_t^2-s\right) \left(12 m_t^4-23 m_t^2 s+8 s^2\right)}{\left(m_t^2-s\right) \left(9 m_t^4-16 m_t^2 s+4 s^2\right)} B_{0(3)}^H
\nonumber\\
& +2\frac{\left(2 m_t^2-3 s\right) \left(4 m_t^2-s\right)}{9 m_t^4-16 m_t^2 s+4 s^2} B_{0(4)}^H
-2 \left(m_t^2-s\right) C_{0(1)}^H
-3 m_t^2 C_{0(2)}^H
\nonumber\\
& +2 \left(4 m_t^2-m_H^2-s\right) C_{0(3)}
\nonumber\\
&+\frac{-67 m_t^6+2 m_t^4 \left(8 m_H^2+67 s\right)-m_t^2 s \left(28 m_H^2+55 s\right)+6 s^2 \left(m_H^2+s\right)}{9 m_t^4-16 m_t^2 s+4 s^2}C_{0(4)}^H
\nonumber\\
& +2 \left(m_t^2-s\right) \left(4 m_t^2-m_H^2-s\right) D_{0(1)}^H \Bigg],
\end{eqnarray}
where the PaVes are analogous to the $Z$ case by replacing $m_Z\to m_H$.

\section{Vanishing of $\hat{d}_t$}
\label{Sec:CEDM}

It is well known that in the SM at the 1-loop level there no exist the CEDM, this statement has come only from the study of the 3-body vertex \cite{Choudhury:2014lna,Aranda:2021,Tavares,Montano-Dominguez:2021eeg}. In principle, the corresponding pseudotensor Dirac structure $\sigma^{\mu\nu}q_\nu\gamma_5$, see Eq.~(\ref{3b-vertex}), could only be induced by the virtual $Z$ and $W$ contributions due to the axial vector term $\gamma^\mu\gamma_5$ present in their respective $Zq\bar{q}$ and $Wq_i\bar{q}_j$ couplings. However, such a structure does not arise at all just after solving their corresponding simple 1-loop tensor integrals.

On the other hand, for the 4-body vertex case, each contributing diagram individually generates the corresponding pseudotensor $\sigma^{\mu\nu}\gamma^5$, see Eqs.~(\ref{4b-vertex}) and (\ref{general-amplitude}), despite this, the whole sum will add up to zero, we show it as follows.

\subsection{$\hat{d}_t(Z)=0$}

For $\hat{d}_t(Z)$ we have a similar expression to that of $\hat{\mu}_t(Z)$ in Eq.~(\ref{Z-CMDM-parts}), but in order to appreciate the cancellation let us apply a tracking tag $L_i=1$ (i=11,12,13,11BS,12BS,13BS) to each subcontribution, this reveals that
\begin{eqnarray}
\hat{d}_t(Z) &= 
 L_{11}\thinspace\hat{d}_{t\thinspace 11}
+L_{12}\thinspace\hat{d}_{t\thinspace 12}
+L_{13}\thinspace\hat{d}_{t\thinspace 13}
\nonumber\\
&
+L_{12\mathrm{BS}}\thinspace\hat{d}_{t\thinspace 11\mathrm{BS}}
+L_{13\mathrm{BS}}\thinspace\hat{d}_{t\thinspace 12\mathrm{BS}}
+L_{14\mathrm{BS}}\thinspace\hat{d}_{t\thinspace 13\mathrm{BS}}
\nonumber\\
&=\frac{i\alpha g_{Vt}g_{At}}{8 \pi c_W^2}
\nonumber\\
&\times \Big[
(L_{11}-L_{12\mathrm{BS}}) 
\Big(
 f_{b1}^Z B_{0(1)}^Z
+f_{b3}^Z B_{0(3)}^Z
+f_{b4}^Z B_{0(4)}^Z
+f_{c4}^Z C_{0(4)}^Z\Big)
\nonumber\\
&+(L_{11\mathrm{BS}}-L_{12})
\Big(
g_{b1}^Z B_{0(1)}^Z+g_{b3}^Z B_{0(3)}^Z
+ g_{c2}^Z C_{0(2)}^Z+g^Z
\Big)
\nonumber\\
&+
(L_{13}-L_{13\mathrm{BS}})
\Big(
 h_{c1}^Z C_{01}^Z
+h_{c2}^Z C_{0(2)}^Z
+h_{c3}^Z C_{0(3)}^Z
+h_{c4}^Z C_{0(4)}^Z
+h_{d1}^Z D_{0(1)}^Z\Big)\Big]
\nonumber\\
&= 0, 
\end{eqnarray}
where $L_{11}-L_{12\mathrm{BS}}$ $=$ $L_{11\mathrm{BS}}-L_{12}$ $=$ $L_{13}-L_{13\mathrm{BS}}$ $=$ 0, the PaVes are given in Sec. \ref{Z-contribution}, and $f_k^Z$, $g_k^Z$ and $h_k^Z$ are functions depending on $m_Z$, $m_t$ and $s$, it is not necessary to show them explicitly. Each $\hat{d}_{t\thinspace i}^Z$ is free of divergences.

\subsection{$\hat{d}_t(W)=0$}

For $\hat{d}_t(W)$ occurs an analogous case to that of $\hat{d}_t(Z)$, it arises from the loop diagrams $(14)-(16)$ in Fig. \ref{FIGURE-Diagrams-4B}. By applying a tracking tag $L_i=1$ (i=14,15,16,14BS,15BS,16BS) to each diagram we obtain
\begin{eqnarray}
\hat{d}_t(W) &= 
 L_{14}\thinspace\hat{d}_{t\thinspace 14}
+L_{15}\thinspace\hat{d}_{t\thinspace 15}
+L_{16}\thinspace\hat{d}_{t\thinspace 16}
\nonumber\\
&
+L_{14\mathrm{BS}}\thinspace\hat{d}_{t\thinspace 14\mathrm{BS}}
+L_{15\mathrm{BS}}\thinspace\hat{d}_{t\thinspace 15\mathrm{BS}}
+L_{16\mathrm{BS}}\thinspace\hat{d}_{t\thinspace 16\mathrm{BS}}
\nonumber\\
&= \frac{i\alpha}{32 \pi}\frac{m_t^2}{m_W^2}\sum_{j=1}^3 \|V_{tq_j}\|^2 
\nonumber\\
&\times
\Big[
(L_{14}-L_{15\mathrm{BS}})
\Big(
 f_{b1}^W B_{0(1)}^W
+f_{b3}^W B_{0(3)}^W
+f_{b4}^W B_{0(4)}^W
+f_{c4}^W C_{0(4)}^W \Big)
\nonumber\\
&+(L_{14\mathrm{BS}}-L_{15})
\Big(
 g_{b1}^W B_{0(1)}^W
+g_{b3}^W B_{0(3)}^W
+g_{c2}^W C_{0(2)}^W \Big)
\nonumber\\
&+ (L_{16}-L_{16\mathrm{BS}})
\Big(
 h_{c1}^W C_{0(1)}^W
+h_{c2}^W C_{0(2)}^W
+h_{c3}^W C_{0(3)}^W
+h_{c4}^W C_{0(4)}^W
+h_{d1}^W D_{0(1)}^W \Big)
\Big]
\nonumber\\
&= 0,
\end{eqnarray}
where $L_{14}-L_{15\mathrm{BS}}$ $=$ $L_{14\mathrm{BS}}-L_{15}$ $=$ $L_{16}-L_{16\mathrm{BS}}$ $=$ 0, the PaVes are given in Sec. \ref{W-contribution}, and $f_k^W$, $g_k^W$ and $h_k^W$ are functions that depend on $m_W$, $m_t$, and $s$. Each $\hat{d}_{t\thinspace i}^W$ is free of divergences.

\section{Results}
\label{Sec:results}

In the following, in order to study the phenomenology of the top quark anomalous CMDM arising from the non-Abelian 4-body vertex function $ggq\bar{q}$, we are going to compare it with that of the Abelian 3-body vertex function $gq\bar{q}$ \cite{Aranda:2021}. For this purpose, we will refer to the anomaly derived from the 4-body vertex as $\hat{\mu}_t^\mathrm{4b}(s)$, whilst the 3-body one will be denoted as $\hat{\mu}_t^\mathrm{3b}(s)$. They depend on the Lorentz-invariant momentum transfer $s$, and will be evaluated at both spacelike ($s=-E^2$) and timelike ($s=E^2$) domains within the energy scale range $E=[10,1000]$ GeV. In advance, the spacelike domain $\hat{\mu}_t^\mathrm{4b}(s)$ will be well behaved in the considered energy range, and will also coincide with the measurement of the experimental central value just at $E=m_Z$. Conversely, in the timelike domain $\hat{\mu}_t^\mathrm{4b}(s)$ will present a very irregular behavior.

As it is known, for studying high-energy physics phenomena in the SM it is conventionally used $\alpha_s(m_Z^2)$~\cite{Workman:2022ynf,Field:1989uq,Deur:2016tte,dEnterria:2022hzv,Yndurain:2006amm,Nesterenko:2016pmx}, conceived in the spacelike regime, same for $s_W(m_Z^2)$, that is to say, the perturbative QCD behavior is counted at the $Z$ gauge boson mass scale. In this sense, $\hat{\mu}_t(s)$ is a quantity obtained in the context of perturbative quantum field theory, for which by means of the $\overline{\mathrm{MS}}$ scheme (see \ref{Appendix-input-values}) we will consider its energy-scale evolution, specifically for the running of $\alpha_s$, also for the small-quark masses $m_{d,s,b}$ involved in the $W$ gauge boson contributing diagrams, and for the top quark mass when $E>2 m_t$ \cite{Barger:1997nn}. In order to achieve these goals, we use the \texttt{RunDec} package~\cite{Chetyrkin:2000yt,Herren:2017osy}; more details can be consulted in \ref{Appendix-input-values}.

Below, the evaluation of the CMDM is presented, and it is worth to keep in mind that, as commented in the Introduction, our derivation of $\hat{\mu}_t^\mathrm{4b}(s)$ is process independent.

\begin{center}
\begin{figure}[!t]
\includegraphics[width=14.5cm]{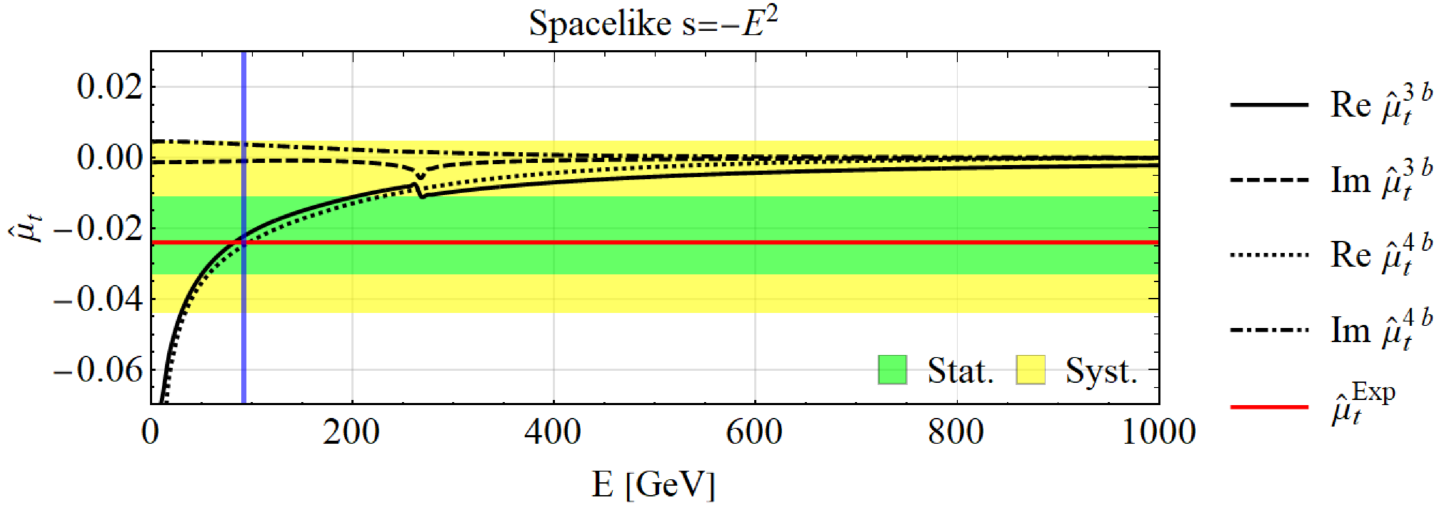}
\caption{Top quark anomalous CMDM in the spacelike domain from the 3- and 4-body vertices. The blue vertical line indicates $E=m_Z$, the experimental value is $\hat{\mu}_t^\mathrm{Exp}$ $=$ $-0.024_{-0.009}^{+0.013}(\mathrm{stat})_{-0.011}^{+0.016}(\mathrm{syst})$.}
\label{FIGURE-CMDM-Esp}
\end{figure}
\end{center}

\begin{table}[!h]
  \centering
\begin{tabular}{|c|c|c|}\hline
\multirow{2}{*}{Contribution} & \multicolumn{2}{c|}{$\hat{\mu}_t(-m_Z^2)$} \\
  \cline{2-3}
         & 3-body        & 4-body      \\
\hline
$g$      & $-2.27\times10^{-2}$ & $-2.54\times10^{-2}$  \\
$\gamma$ & $2.62\times10^{-4}$  & $-5.19\times10^{-4}$ \\
$Z$      & $-1.76\times10^{-3}$ & $-1.78\times10^{-3}$ \\
$W$      & $-2.89\times10^{-5}-9.23\times10^{-4}i$ & $-3.43\times10^{-4}+3.84\times10^{-3}i$ \\
$H$      & $1.85\times10^{-3}$  & $3.06\times10^{-3}$ \\
Total    & $-2.24\times10^{-2}-9.23\times10^{-4}i$ & $-2.5\times10^{-2}+3.84\times10^{-3}i$ \\
\hline
\end{tabular}
\caption*{(a)}
\begin{tabular}{|c|c|c|}\hline
\multirow{2}{*}{Contribution} & \multicolumn{2}{c|}{$\hat{\mu}_t(-m_Z^2)$} \\
  \cline{2-3}
         & 3-body             & 4-body     \\
\hline
QCD      & $-2.27\times10^{-2}$   & $-2.54\times10^{-2}$ \\
EW       & $-1.52\times10^{-3}-9.23\times10^{-4}i$ & $-2.64\times10^{-3}+3.84\times10^{-3}i$ \\
YK       & $1.85\times10^{-3}$    & $3.06\times10^{-3}$ \\
Total    & $-2.24\times10^{-2}-9.23\times10^{-4}i$ & $-2.50\times10^{-2}+3.84\times10^{-3}i$ \\
\hline
\end{tabular}
\caption*{(b)}
\caption{Top quark anomalous CMDM in the spacelike domain at $s=-m_Z^2$:
(a) individual particle contributions, (b) sector contributions.
The total absolute values are
$\|\hat{\mu}_t^\mathrm{3b}\|=0.0224$
and
$\|\hat{\mu}_t^\mathrm{4b}\|=0.0253$.
The experimental central value is $\hat{\mu}_t^\mathrm{Exp}=-0.024$ \cite{Sirunyan:2019eyu}.}
\label{TABLE-ACMDM-Esp}
\end{table}

\subsection{The spacelike evaluation $s=-E^2$}

In Fig.~\ref{FIGURE-CMDM-Esp} the behavior of $\hat{\mu}_t^\mathrm{3b}(s)$ and $\hat{\mu}_t^\mathrm{4b}(s)$ are plotted for the spacelike domain, $s=-E^2$, considering $E=[10,1000]$ GeV, where the red line corresponds to the central value of the experimental measurement $\hat{\mu}_t^\mathrm{Exp}$ $=$ $-0.024$, and the blue line indicates $E=m_Z$. The real part curves are extremely close to each other, essentially the same, in the range $E=[10,290]$ GeV; after 290 GeV both signals separate slightly. On the contrary, their imaginary parts have different signs.

In Table~\ref{TABLE-ACMDM-Esp} the evaluation for $\hat{\mu}_t(-m_Z^2)$ is listed. The Table~\ref{TABLE-ACMDM-Esp}(a) represents the details of each virtual particle contribution: the gluon part provides the largest values for both the three- and 4-body cases, while the photon contribution yields the smallest ones. It is important to highlight that, once again, the imaginary part comes strictly from $\hat{\mu}_t^\mathrm{4b}(W)$, just as occurred in the 3-body case $\hat{\mu}_t^\mathrm{3b}(W)$. It is remarkable that Re$\thinspace\hat{\mu}_t^\mathrm{4b}(-m_Z^2)=$ $-0.025$ is even closer to $\thinspace\hat{\mu}_t^\mathrm{Exp}=-0.024$ than Re$\thinspace\hat{\mu}_t^\mathrm{3b}(-m_Z^2)=$ $-0.0224$. On the other hand, by comparison of the absolute values
$\|\hat{\mu}_t^\mathrm{3b}\|=$ $0.0224$
and
$\|\hat{\mu}_t^\mathrm{4b}\|=$ $0.0253$, both results are compatible with the experimental central value $\|\hat{\mu}_t^\mathrm{Exp}\|=0.024$.

\subsection{The timelike evaluation $s=E^2$}

\begin{center}
\begin{figure}[!t]
\includegraphics[width=14.5cm]{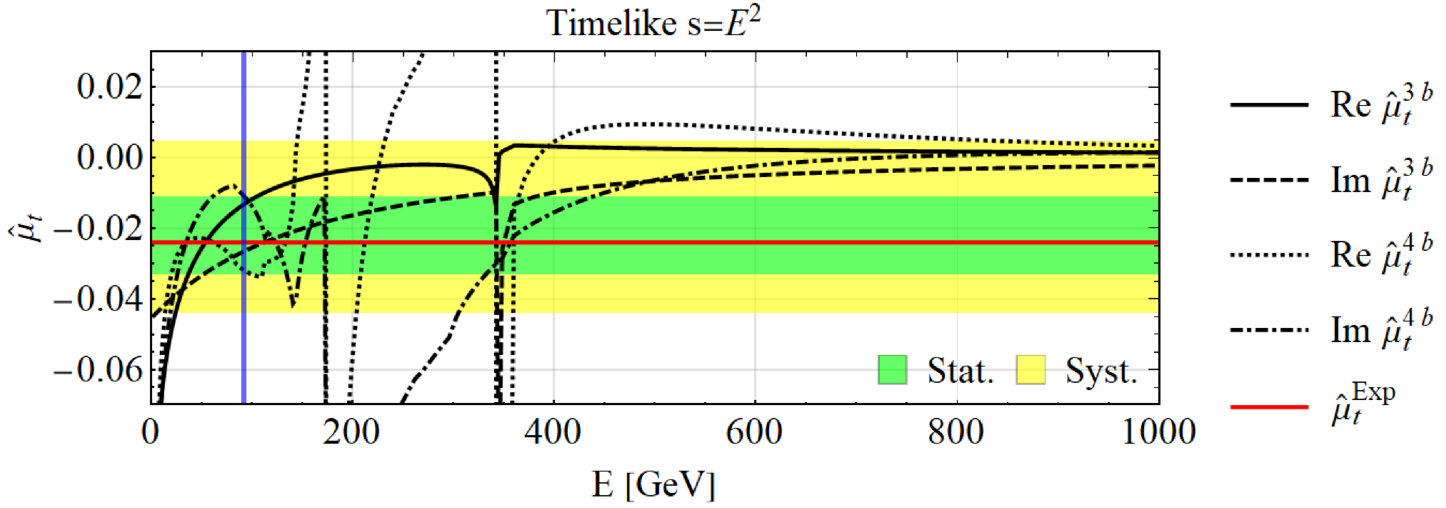}
\caption{Top quark anomalous CMDM in the timelike domain from the 3- and 4-body vertices. The blue vertical line indicates $E=m_Z$, and the experimental value is $\hat{\mu}_t^\mathrm{Exp}$ $=$ $-0.024_{-0.009}^{+0.013}(\mathrm{stat})_{-0.011}^{+0.016}(\mathrm{syst})$.}
\label{FIGURE-ACMDM-Temp}
\end{figure}
\end{center}

Concerning the evaluation of the timelike domain, as it can be observed in Fig.~\ref{FIGURE-ACMDM-Temp}, $\hat{\mu}_t^\mathrm{4b}(s)$ presents an irregular behavior. It is worth to mention that this type of irregular behavior also appears in both the fine structure constant \cite{Jegerlehner:2011mw} (it also develops an imaginary part) and the weak mixing angle \cite{Jegerlehner:2017zsb}. Such irregular behaviour is shown in Fig.~\ref{FIGURE-ACMDM-Temp}, where $\hat{\mu}_t^\mathrm{3b}(s)$ has a discontinuity at the threshold $E=2m_t$, while $\hat{\mu}_t^\mathrm{4b}(s)$ has discontinuities at $E=m_t$ and the threshold. Specifically, the particular evaluation at $E=m_Z$ is listed in Table~\ref{TABLE-ACMDM-Temp}. It should be noted that the gluon contribution contains an imaginary part, which does not occur in the spacelike domain, where the imaginary part only comes from the $W$ gauge boson contribution. By contrasting their respective absolute values
$\|\hat{\mu}_t^{\mathrm{3b}}\|$  $=$ $0.0298$
and
$\|\hat{\mu}_t^{\mathrm{4b}}\|$  $=$ $0.0335$, this indicates that the timelike domain results disfavored with respect to the experimental central value $\|\hat{\mu}_t^\mathrm{Exp}\|=0.024$.

Our findings for $\hat{\mu}_t^\mathrm{4b}(s)$ repeat the same features of $\hat{\mu}_t^\mathrm{3b}(s)$: i) it is IR divergent when the Lorentz-invariant momentum transfer of the vertex vanishes, $s=0$; ii) the Re$\thinspace\hat{\mu}_t^\mathrm{4b}(s)$ offers very close that Re$\thinspace\hat{\mu}_t^\mathrm{3b}(s)$ when a spacelike domain is considered, providing for $E=m_Z$ an even closer result to the experimental one; iii) the spacelike Im$\thinspace\hat{\mu}_t^\mathrm{4b}(s)$ part arises strictly from the virtual $W$ boson contribution, just as happen for $\thinspace\hat{\mu}_t^\mathrm{3b}(s)$; iv) the imaginary part of the timelike evaluation also receives contribution from the $W$ diagrams as well as from diagrams with virtual gluons; v) finally, the timelike evaluation at $s=m_Z^2$, just as it occurs with $\hat{\mu}_t^\mathrm{3b}(s)$, is also disadvantaged with respect to the experimental result.

\newpage

\begin{table}[!h]
  \centering
\begin{tabular}{|c|c|c|}\hline
\multirow{2}{*}{Contribution} & \multicolumn{2}{c|}{$\hat{\mu}_t(m_Z^2)$} \\
  \cline{2-3}
         & 3-body        & 4-body     \\
\hline
$g$      & $-1.38\times10^{-2}-2.55\times10^{-2}i$ & $-2.62\times10^{-2}-1.44\times10^{-2}i$ \\
$\gamma$ & $2.88\times10^{-4}$  & $-6.76\times10^{-4}$ \\
$Z$      & $-1.88\times10^{-3}$ & $-1.85\times10^{-3}$ \\
$W$      & $1.41\times10^{-4}-1.16\times10^{-3}i$ & $-6.24\times10^{-3}+3.78\times10^{-3}i$ \\
$H$      & $1.98\times10^{-3}$ & $3.19\times10^{-3}$ \\
Total    & $-1.33\times10^{-2}-2.67\times10^{-2}i$ & $-3.18\times10^{-2}-1.06\times10^{-2}i$ \\
\hline
\end{tabular}
\caption*{(a)}
\begin{tabular}{|c|c|c|}\hline
\multirow{2}{*}{Contribution} & \multicolumn{2}{c|}{$\hat{\mu}_t(m_Z^2)$} \\
  \cline{2-3}
         & 3-body           & 4-body      \\
\hline
QCD      & $-1.38\times10^{-2}-2.55\times10^{-2}i$ & $-2.62\times10^{-2}-1.44\times10^{-2}i$ \\
EW       & $-1.45\times10^{-3}-1.16\times10^{-3}i$ & $-8.77\times10^{-3}+3.78\times10^{-3}i$ \\
YK       & $1.98\times10^{-3}$ & $3.19\times10^{-3}$ \\
Total    & $-1.33\times10^{-2}-2.67\times10^{-2}i$ & $-3.18\times10^{-2}-1.06\times10^{-2}i$ \\
\hline
\end{tabular}
\caption*{(b)}
\caption{Top quark anomalous CMDM in the timelike domain at $s=m_Z^2$: (a) individual particle contributions, (b) sector contributions.
The total absolute values are
$\|\hat{\mu}_t^{\mathrm{3b}}\|$  $=$ $0.0298$
and
$\|\hat{\mu}_t^{\mathrm{4b}}\|$  $=$ $0.0335$.
The experimental central value is $\hat{\mu}_t^\mathrm{Exp}=-0.024$ \cite{Sirunyan:2019eyu}.}
\label{TABLE-ACMDM-Temp}
\end{table}

\section{Conclusions}
\label{Sec:conclusions}

We have presented a new approach to address the calculation of the anomalous CMDM of the top quark in the SM, where the contributions to this observable are extracted via the radiative correction at the 1-loop level of the 4-body vertex function $ggt\bar{t}$, in a process independent way. This alternative scheme comprises more than seventy Feynman diagrams of triangle and box topologies, which exclusively arises from the non-Abelianity of QCD. Our most important prediction comes from the high-energy scale evaluation of the top quark CMDM just at the $Z$ gauge boson mass.

The top quark CMDM evaluated in the spacelike domain $s=-E^2$, where $E=[10,1000]$ GeV, is well behaved, in contrast with the timelike domain $s=E^2$. In particular, right at the energy scale of the $Z$ gauge boson mass, our prediction is $\hat{\mu}_t^\mathrm{4b}(-m_Z^2)=$ $-0.025+0.00384i$, whose real part is even closer to the experimental central value $\hat{\mu}_t^\mathrm{Exp}=$ $-0.024$ than that coming from the 3-body vertex function. Conversely, the timelike result is $\hat{\mu}_t^\mathrm{4b}(m_Z^2)=$ $-3.18\times10^{-2}-1.06\times10^{-2}i$. Based on the literature and our results, it seems to be that the postulation of a spacelike behaviour of the CMDM of the top quark is consistent with the current experimental result.

The consequences of our $\hat{\mu}_t$ could be very exciting indeed, because this quantity certainly is not small. In fact, we could say that to be a quantity from a quantum fluctuation it has turned out to be a large value. Thus, from already known tree-level dispersion or decay processes, we could test the impact of our chromodipolar couplings $gt\bar{t}$ and $ggt\bar{t}$ as effective pieces inserted in those processes and predict possible deviations that might be tested experimentally in the physics of the top quark at the LHC.

\section*{Acknowledgments}
This work has been partially supported by SNI-CONAHCYT and CIC-UMSNH. JMD thanks to the CONAHCYT program Investigadoras e Investigadores por M\'exico, project 1753. We thank the three referees for their constructive comments during the review of this manuscript.

\appendix

\section{Infrared divergent scalar functions}
\label{Appendix:IR-PaVes}

The IR divergent PaVes that are present in the $\hat{\mu}_t^{\mathrm{4b}}(g)$ contribution, obtained with $\texttt{Package-X}$, are given by
\begin{eqnarray}\label{C01IR}
C_{0(1)}^g &\equiv C_0(0,s,-2m_{q}^2+3s\thinspace;0,0,0)
\nonumber \\
&= \frac{1}{4\left(m_{q}^2-s\right)}
\left(
2\Delta_\mathrm{IR}+\ln\frac{\mu^2}{2m_q^2-3s}+\ln\frac{\mu^2}{-s}
\right)\ln\frac{2m_q^2-3s}{-s},
\end{eqnarray}
\begin{eqnarray}\label{C04IR}
C_{0(4)}^g &\equiv C_0(m_{q_i}^2,0,s\thinspace;m_{q},0,0)
\nonumber\\
&= \frac{-1}{2(m_{q}^2 - s)} \Bigg[\Delta_\mathrm{IR2}+\Delta_\mathrm{IR}\left(\ln\frac{\mu ^2}{m_q^2}+2\ln\frac{m_q^2}{m_q^2-s}\right)
+2\ln\frac{\mu ^2}{m_q^2}\ln\frac{m_q^2}{m_q^2-s}
\nonumber\\
&+\frac{1}{2}\ln^2\frac{\mu ^2}{m_q^2}
-2 \mathrm{Li}_2\frac{s}{s-m_q^2}
+\ln^2\frac{m_q^2}{m_q^2-s}
+\frac{\pi ^2}{12} \Bigg],
\end{eqnarray}
\begin{eqnarray}\label{C06IR}
C_{0(6)}^g &\equiv C_0(m_{q}^2,m_{q}^2,s\thinspace;m_{q},0,m_{q})
\nonumber\\
&= \frac{-1}{4 m_{q}^2-s}
\left(\Delta_\mathrm{IR}+\ln\frac{\mu ^2}{m_{q}^2}\right)
\frac{R_1}{s} \ln \frac{2 m_q^2-s+R_1}{2m_q^2}
-\frac{1}{R_1}
\bigg[
2\mathrm{Li}_2\frac{2 m_q^2-s+R_1}{-2 m_q^2}
\nonumber\\
&-\frac{1}{2}\ln\frac{2 m_q^2-s+R_1}{2 m_q^2}
\left(
\ln\frac{m_q^2}{s-4 m_q^2}+\ln\frac{2 m_q^2-s-R_1}{2(s-4 m_q^2)}
\right)
+\frac{\pi^2}{6}
\bigg],
\end{eqnarray}
\begin{eqnarray}\label{D01IR}
D_{0(1)}^g &\equiv D_0(m_{q}^2,m_{q}^2,0,-2m_{q}^2+3s,s,s\thinspace;0,m_{q},0,0)
\nonumber\\
&= \frac{1}{s\left(s-m_{q}^2\right)}
\Bigg\{
\frac{\Delta_\mathrm{IR2}}{2}
+\Delta_\mathrm{IR}\left(\frac{1}{2}\ln\frac{\mu^2}{m_q^2}+\ln\frac{m_q^2}{m_q^2-s}+\ln\frac{2m_q^2-3s}{-s}\right)
\nonumber\\
&+\ln\frac{\mu^2}{m_q^2}\left(\ln\frac{m_q^2}{m_q^2-s}+\ln\frac{2m_q^2-3s}{-s}\right)
+\frac{1}{4}\ln^2\frac{\mu^2}{m_q^2}
-2\mathrm{Li}_2\frac{2 m_q^2-2s}{s}
\nonumber\\
&+\ln^2\frac{m_q^2}{m_q^2-s}
+2\ln\frac{m_q^2}{m_q^2-s}\ln\frac{2 m_q^2-3s}{-s}-\frac{\pi ^2}{8}
\Bigg\},
\end{eqnarray}
\begin{eqnarray}\label{D02IR}
D_{0(2)}^g &\equiv D_0(m_{q}^2,m_{q}^2,0,-2m_{q}^2+3s,s,s\thinspace;m_{q},0,m_{q},m_{q})
\nonumber\\
&= \frac{1}{\left(m_q^2-s\right)\left(4 m_q^2-s\right)}\left(\Delta_\mathrm{IR}+\ln\frac{\mu ^2}{m_q^2}\right)
\frac{R_1}{s} \ln \frac{2 m_q^2-s+R_1}{2m_q^2}
\nonumber\\
& +\frac{2}{\left(m_q^2-s\right)R_1}\Bigg[
\frac{R_1}{4 m_q^2-s}
\left(\ln\frac{m_q^2}{m_q^2-s}-\ln\frac{4 s R_1}{(s+R_1)^2}\right)
\frac{R_1}{s} \ln \frac{2 m_q^2-s+R_1}{2m_q^2}
\nonumber\\
&
+\frac{1}{2}\ln^2\frac{4 m_q^2-3 s+R_2}{2 m_q^2}
+\frac{1}{2}\mathrm{Li}_2\left(\frac{s-R_1}{s+R_1}\right)^2-\frac{\pi^2}{12}
\nonumber\\
&
+\mathcal{L}i_2\left(
\frac{R_1-s}{R_1+s}-i \varepsilon,
\frac{4 m_q^2-3 s+R_2}{2 m_q^2}
+\frac{R_2}{m_q^2}i\varepsilon
\right)
\nonumber\\
&+\mathcal{L}i_2\left(
\frac{R_1-s}{R_1+s}-i\varepsilon,\frac{4 m_q^2-3 s-R_2}{2 m_q^2}
-\frac{R_2}{m_q^2}i\varepsilon
\right)
\Bigg],
\end{eqnarray}
where $R_1\equiv\sqrt{s(s-4m_q^2)}$, $R_2\equiv$ $\sqrt{3(2m_q^2-3s)(2m_q^2-s)}$, $\mathcal{L}i_2$ is the Beenakker-Denner continued dilogarithm and the IR poles are
\begin{eqnarray}\label{}
\Delta_\mathrm{IR} &\equiv \frac{(4\pi)^{\epsilon_\mathrm{IR}}\Gamma(1+\epsilon_\mathrm{IR})}{\epsilon_\mathrm{IR}}
=
(4\pi)^{\epsilon_\mathrm{IR}}\Gamma(\epsilon_\mathrm{IR})
\nonumber\\
&\approx \frac{1}{\epsilon_\mathrm{IR}}
-\gamma_E+\ln 4\pi,
\end{eqnarray}
\begin{eqnarray}\label{}
\Delta_\mathrm{IR2} &\equiv \frac{(4\pi)^{\epsilon_\mathrm{IR}}\Gamma(1+\epsilon_\mathrm{IR})}{\epsilon_\mathrm{IR}^2}
= \frac{(4\pi)^{\epsilon_\mathrm{IR}}\Gamma(\epsilon_\mathrm{IR})}{\epsilon_\mathrm{IR}}
\nonumber\\
&\approx
\frac{1}{\epsilon_\mathrm{IR}^2}+\frac{1}{\epsilon_\mathrm{IR}}\left(-\gamma_E+\ln4\pi\right)
+\frac{\gamma_E^2}{2}-\gamma_E\ln4\pi
+\frac{1}{2}\ln^2 4\pi+\frac{\pi^2}{12}.
\end{eqnarray}

\section{Input values}
\label{Appendix-input-values}

In our calculations we use the electron unit charge $e=\sqrt{4\pi\alpha}$ and the QCD group strong coupling constant $g_s=\sqrt{4\pi\alpha_s}$. We consider input values from PDG 2020 and 2021 update \cite{Workman:2022ynf}: the strong coupling constant $\alpha_s(m_Z^2)=0.1179$, the weak-mixing angle $s_W\equiv$ $\sin{\theta_W}(m_Z^2)$ $=$ $\sqrt{0.23121}$, the boson masses $m_W$=$80.379$ GeV, $m_Z$=$91.1876$ GeV and $m_H=125.25$ GeV. The Cabibbo-Kobayashi-Maskawa matrix is
\begin{eqnarray}\label{}
V_\mathrm{CKM} &=&
\left(
\begin{array}{ccc}
|V_{ud}| & |V_{us}| & |V_{ub}| \\
|V_{cd}| & |V_{cs}| & |V_{cb}| \\
|V_{td}| & |V_{ts}| & |V_{tb}| \\
\end{array}
\right)
=
\left(
\begin{array}{ccc}
 0.9737 & 0.2245 & 0.00382 \\
 0.221 & 0.987 & 0.041 \\
 0.008 & 0.0388 & 1.013 \\
\end{array}
\right);
\end{eqnarray}
the electric charges of the quarks $Q_{t}=2/3$ and the weak couplings $g_V^t=1/2-4/3\thinspace s_W^2$ and $g_A^t=1/2$. The fine-structure constant $\alpha(m_Z^2)=1/129$ is taken from \cite{Denner:2019vbn}.

We apply the $\overline{\mathrm{MS}}$ scheme by means of $\texttt{RunDec}$ \cite{Chetyrkin:2000yt,Herren:2017osy} to consider the energy scale evolution of $\alpha_s$ and the quark masses. Since the $\hat{\mu}_t(W)$ contribution participate the $d$, $s$ and $b$ quarks, we firstly resort to their masses provided by PDG and then take them to the desired energy scale. Thus, in the $\overline{\mathrm{MS}}$ scheme, at low energy scale, we have
$m_d(2~\mathrm{GeV}) = 0.00467$  GeV,
$m_s(2~\mathrm{GeV}) = 0.093$  GeV,
$m_b(m_b) = 4.18$ GeV,
and the top quark $m_t = 172.5$ GeV is the pole mass. At high energy scale, for example, at the scale of the $Z$ gauge boson mass, these masses are
$m_d(m_Z)=0.00266$ GeV,
$m_s(m_Z)=0.0530$ GeV and
$m_b(m_Z)=2.863$ GeV, with five active quarks.
For the top quark case, we consider its mass evolution from $m_t(E>2m_t)$ with six active quarks, for example, $m_t(1000~\mathrm{GeV})=152.90$ GeV when $\alpha_s(1000~\mathrm{GeV})=$ $0.0884$.

\end{document}